\documentclass[twocolumn,amssymb,nobibnotes,aps,prl,showpacs]{revtex4}
  \usepackage[dvips]{color}
  \usepackage{newlfont}
  \usepackage{graphicx}
  \usepackage{amssymb}
  \usepackage{amsmath}
  \usepackage[latin1]{inputenc}
  \usepackage{float}

\begin{document}
\title{Two-level quantum dynamics, integrability and unitary NOT gates}
\author{Renato M. Angelo}
\affiliation{Universidade Federal do Paraná, Departamento de Física, Caixa Postal 19044, Curitiba 81531 990, PR, Brazil.}
%\email{renato@fisica.ufpr.br}
\author{Walter F. Wreszinski}
%\email{wreszins@fma.if.usp.br}
\affiliation{Universidade de São Paulo, Instituto de Física,\\ 
Caixa Postal 66318, São Paulo 05315 970, SP, Brazil.}
%\date{February 15, 2005}

\begin{abstract}
We study the dynamics of a two-level quantum system interacting with an external electromagnetic field periodic and quasiperiodic in time. The quantum evolution is described exactly by the classical equations of motion of a gyromagnet in a time-dependent magnetic field. We prove that this classical system is integrable as a consequence of the underlying unitary quantum dynamics.  As a consequence, for the periodic case: i) rigorous assessment of the validity of the rotating-wave approximation (RWA) becomes possible even beyond the assumptions of resonance and weak coupling (the latter conditions are also shown to follow from the method of averaging); ii) we determine conditions for the realization of the quantum NOT operation beyond the RWA, by means of classical stroboscopic maps. The results bear upon areas as diverse as quantum optics, nuclear magnetic resonance (NMR) and quantum computation.


\end{abstract}

\maketitle

%===========================================================================
%===========================================================================
Two-level systems have achieved a paradigmatic place in modern quantum mechanics with the advent of strong laser pulses and advanced NMR techniques. By coupling a two-level system to a time-dependent classical field it is possible to describe the equivalent dynamics of a two-level atom interacting with an external electromagnetic field (quasi)periodic in time in the dipole approximation (relevant to quantum optics) or of a nuclear spin $1/2$ in an external (quasi)periodic magnetic field (relevant to NMR and to the NOT operation in quantum computation). They are modelled by the same Hamiltonian
%--------------
\begin{eqnarray}
H(t)=-\frac{1}{2}\mathbf{B}(t)\cdot\mathbf{\Sigma}, \label{HQ}
\end{eqnarray}
%--------------
where $\mathbf{\Sigma}=(\sigma_x,\sigma_y,\sigma_z)$ is the vector composed by the Pauli matrices. For simplicity we adopted $\hbar=1$ and thus the magnetic field is in units of frequency. We shall analyze two important external periodic fields frequently used in quantum optics and NMR experiments, particularly for the implementation of the quantum NOT gate. The first one is a radio-frequency field \cite{vander04} given by
%--------------
\begin{eqnarray}
\mathbf{B}_R=-2(B_0 \cos{\omega t}, B_0 \sin{\omega t}, B_1), \label{BR}
\end{eqnarray}
%--------------
where $B_0$ and $B_1$ are constant amplitudes. We call it {\em rotating} (R) because it corresponds to a magnetic field which rotates around the $z$-axis. In quantum optics it corresponds to a circularly polarized radiation field. The dynamics for this field may be solved analytically by applying the unitary transformation $\exp(-\imath\, \omega t\, \sigma_z)$ to the wave-function in Schrödinger equation for the Hamiltonian \eqref{HQ}; in the resulting rotating frame the Hamiltonian becomes time-independent and the solution emerges easily \cite{baym}. We also investigate the dynamics under a periodic {\em nonrotating} (NR) magnetic field along a preferred direction:
%--------------
\begin{eqnarray}
\mathbf{B}_{NR}=-2(B_2\cos{\omega t},0,B_3), \label{BNR}
\end{eqnarray}
%--------------
with $B_2$ and $B_3$ time-independent amplitudes. $\mathbf{B}_{NR}$ corresponds to a plane-polarized radiation field in quantum optics \cite{nus73}, a fact which explains its conceptual importance. In fact, the quantum dynamics corresponding to $\mathbf{B}_{NR}$ is very rich and far from soluble: it has been a recurrent theme in quantum optics since the seminal papers of Bloch and Siegert \cite{bloch} and Autler and Townes \cite{autler}. This fact has often led to the replacement of $\mathbf{B}_{NR}$ by $\mathbf{B}_{R}$, in particular for the dynamics of the quantum computation \cite{divicenzo95}. However, the R field is much more difficult to manufacture and, therefore, the most commonly used in NMR experiments, in particular for quantum computational purposes, is $\mathbf{B}_{NR}$ \cite{divicenzo95}. Thus, in each of the above areas, quantum optics, NMR and quantum computation, it is of central importance to understand the dynamics induced by $\mathbf{B}_{NR}$.

The above-mentioned replacement of $\mathbf{B}_{NR}$ by $\mathbf{B}_{R}$ is called rotating-wave approximation (RWA). It has been heuristically discussed in many text books (see \cite{baym} for the most careful presentation), but we do not know any exact statement on the domain of validity of the various parameters (in contrast, an exact method, which may be called rotating-wave transformation, has been introduced in \cite{jauslin}). Such a statement may be obtained from the following result.

\vspace{0.2cm}
\noindent
\underline{Result 1}: Let $B_2 T\ll 1$ (weak coupling), with $T=2\pi/\omega$. Then, if $\omega=2 B_3$ (resonance), the solution of the Schrödinger equation $\psi_R$ corresponding to $H(t)$ given by \eqref{HQ} with $\mathbf{B}=\mathbf{B}_{R}$ (with $B_0=B_2/2$ and $B_1=B_3$) approximates the solution $\psi_{NR}$ of the Schrödinger equation with the Hamiltonian \eqref{HQ} with $\mathbf{B}=\mathbf{B}_{NR}$ in the sense that $||\psi_{R}(t)-\psi_{NR}(t) ||=\cal{O}(B_2 T)$, whenever $0\le t \le \cal{O}(\frac{1}{B_2 T})$.

\vspace{0.2cm}
Above, $||\cdot||$ is the Euclidean norm in the spinor vector space. Writing
%--------------
\begin{eqnarray}
\psi(t)=\exp\left(- \frac{\imath\,\omega\, t\,\sigma_z}{2}\right)\psi'(t)
\label{transf}
\end{eqnarray}
%--------------
we obtain for $\psi'$ under the condition of resonance the Schrödinger equation
\begin{subequations}
%--------------
\begin{eqnarray}
\imath \frac{d\psi'}{dt}=\epsilon f(\psi',t)
\end{eqnarray}
%--------------
where $\epsilon=B_2\, \omega^{-1}$ is, by assumption, a small parameter, and 
%--------------
\begin{eqnarray}
f(\psi',t)=\left[\frac{\omega\, \sigma_x}{2}+\frac{\omega}{2}\left(\sigma_x \cos{2\omega t}-\sigma_y \sin{2\omega t} \right) \right]\psi'
\end{eqnarray}
%--------------
\label{psil}
\end{subequations}
is periodic in $t$, of average equal to $\frac{\omega\, \sigma_x}{2}\psi'$. By the averaging theorem (\cite{verhulst90}, Theorem 11.1), the solution $\psi'$ of \eqref{psil} is $\cal{O}(B_2 T)$-close to the solution of 
%--------------
\begin{eqnarray}
\imath \frac{d \psi_0'}{dt}=\left(B_2\,\omega^{-1} \right)\frac{\omega\, \sigma_x}{2}\psi'
\end{eqnarray}
%--------------
if $0\le t\le \cal{O}(\frac{1}{B_2 T})$; undoing the transformation \eqref{transf}, we obtain that $\psi$ is close to the function $\psi_R(t)=e^{-\imath \omega t \sigma_z/2}e^{-\imath B_2 t \sigma_x /2}\psi_R(0)$, which is the wave-function in the RWA.

A different approach has been pioneered by Feynman {\it et al} \cite{feynman} (see also \cite{hioe84} and \cite{bagrov01}). Considering a density operator $\rho$ satisfying the conditions of unitary evolution ($\imath \dot{\rho}=[H(t),\rho]$), normalization ($\textrm{Tr}\rho=1$) and purity ($\rho²=\rho$; this condition is, however, not essential to the method), one may define the parametrization
%--------------
\begin{eqnarray}
\rho=\frac{1}{2}\left(\mathbf{1}+\mathbf{S}\cdot\mathbf{\Sigma} \right) \label{rhoS}
\end{eqnarray}
%--------------
to map the quantum dynamics in terms of classical Hamiltonian formalism \cite{feynman,bagrov01} defined by
%--------------
\begin{subequations}
\begin{eqnarray}
&\cal{H}(t)=-\mathbf{B}(t)\cdot \mathbf{S},& \label{HC} \\
&\displaystyle{\frac{d\mathbf{S}}{dt}}=\mathbf{S}\times\mathbf{B},& \label{edS}\\ &\mathbf{S}²=1.&\label{S2}
\end{eqnarray} \label{HS}
\label{gyro}
\end{subequations}
%--------------
Equations \eqref{gyro} define the classical geometric picture associated to quantum two-level systems: the unit vector $\mathbf{S}$ precesses around the vector $\mathbf{B}$ just like a {\it classical gyromagnet} precesses in a magnetic field. By means of a canonical transformation, $\mathbf{S}=(\sqrt{1-q^2}\cos{p},\sqrt{1-q^2}\sin{p},-q)$, we get the following bi-dimensional canonical representation
%--------------
\begin{subequations}
\begin{eqnarray}
\cal{H}&=&-\left(B_x\cos{p}+B_y \sin{p}\right)\sqrt{1-q^2}-B_z q,\label{HqpA}\\
\dot{q}&=&\left(B_x \sin{p}-B_y\cos{p} \right)\sqrt{1-q²}, \\ 
\dot{p}&=&-\left(B_x\cos{p}+B_y \sin{p}\right) \frac{q}{\sqrt{1-q²}}-B_z,
\end{eqnarray} \label{Hqp}
\end{subequations}
%--------------
in which the amplitudes $B_{x,y,z}$ are functions of time. The square roots have as consequence that \eqref{Hqp} does not satisfy a Lipschitz condition (see, e.g., \cite{verhulst90}, Theorem 1.2) and thus there is not, a priori, a local existence and uniqueness theorem (see, e.g., \cite{verhulst90}, Theorem 1.1). Thus, a comparison with the (nonsingular) dynamics \eqref{edS} is necessary to select the correct solutions. Representation \eqref{Hqp} is, however, important for the construction of the stroboscopic maps, which are numerical tools to attest the integrability of the system.

We have now: 

\vspace{0.2cm}
\noindent
\underline{Result 2}: The system (8) is integrable.\\
In order to show this we consider the classical distance function on the Bloch sphere
%--------------
\begin{subequations}
\begin{eqnarray}
D(t)\equiv ||\,\mathbf{S}_1-\mathbf{S}_2\,||,\label{DS}
\end{eqnarray}
%--------------
where different indices refer to different initial conditions. By \eqref{rhoS}, it follows from $(\mathbf{a}\cdot\mathbf{\Sigma})(\mathbf{b}\cdot\mathbf{\Sigma})=(\mathbf{a}\cdot\mathbf{b})+\imath \mathbf{\Sigma}\cdot(\mathbf{a}\times\mathbf{b})$ that
%--------------
\begin{eqnarray}
D(t)= \sqrt{2\, \textrm{Tr}\left(\rho_1-\rho_2 \right)²}, \label{Drho}
\end{eqnarray}
\end{subequations}
%--------------
where the subindices refer to different initial conditions on the Schrödinger wave-function $\rho_{1,2}(t)=|\psi_{1,2}(t)\rangle\langle\psi_{1,2}(t)|$, with
%--------------
\begin{eqnarray}
\rho_{1,2}(t)= U(t)\,\rho_{1,2}(0)\, U^{\dag}(t), \label{rhoU}
\end{eqnarray}
%--------------
where $U$ is the unitary propagator satisfying $\imath\, \dot{U}=H(t)U(t)$. Equation \eqref{DS} allows us to define a Lyapunov exponent as
%--------------
\begin{eqnarray}
\lambda=\lim\limits_{D(0)\to 0}\lim\limits_{t\to\infty} \frac{1}{t} \ln\left[\frac{D(t)}{D(0)}\right]. \label{lambda}
\end{eqnarray}
%--------------
However, by \eqref{Drho} and \eqref{rhoU} it is immediate that $D(t)=D(0)$ and thus, by \eqref{lambda}, $\lambda=0$. When a full set (i.e., equal to the number of degrees of freedom) of constants of the motion in involution does not exist, so that we have a system which is not integrable, sensivity to initial conditions is certain to exist at least in some region of phase space, i.e., there is inevitably chaotic behavior. Thus we may regard the proof that $\lambda=0$ as a proof of integrability, although it may be difficult to find explicitly an additional constant of the motion for the general autonomous system equivalent to \eqref{HS} (see, e.g., \cite{bagrov01}, for the description of these autonomous systems). It is worth emphasizing that our proof of integrability is valid for an arbitrary time-dependent field, including the quasiperiodic case. This contradicts the basic assertion of \cite{pomeau} and confirms the result of \cite{badii}. More importantly, we show the real reason for the integrability observed in \cite{badii} using double Poincaré sections. 

Equations \eqref{gyro} describe the interaction of one two-level atom with the quantized radiation field (q.r.f.) in the Floquet approximation, where the average number of photons becomes very large, keeping the photon density constant \cite{jauslin97}. It may also be regarded as a special case of the (mean field or classical) limit of the expectation-values of the Heisenberg equations of motion for a system of many atoms interacting with the q.r.f. in suitable (e.g. coherent) states \cite{hepp}. In this limiting process, Result 2 is no longer {\it generally} true and, indeed, in the single-mode case, in which the electric field is periodic in time, the resulting equations do exhibit chaotic behavior for the model with an initial population, when the polarization (i.e. the macroscopic transition dipole moment) is nonzero \cite{milonni}. Only the special situation of zero polarization $p$ reduces to our equations \eqref{gyro}, in which case it is wrongly asserted in \cite{milonni} that the model is integrable due to condition \eqref{S2}. Indeed, \eqref{S2} solely reduces the three spin variables to two independent ones, i.e., one obtains a one-degree of freedom system in an external periodic field. The equivalent autonomous system is thus in general chaotic due to the nonlinearity in the time-variable \cite{bagrov01}. It is our Result 2 which shows that integrability holds as a consequence of the unitarity of the underlying quantum dynamics. Finally, the full one-mode quantum Hamiltonian has striking similarity to our case, because it is unitarily equivalent to a Hamiltonian where the q.r.f. and the atom are decoupled \cite{lo}, and thus is integrable in this sense. The reason of this similarity is that the Floquet approximation preserves some of the quantum structure \cite{jauslin97}, while the classical approximation (limit) is rather singular \cite{hepp}. Only in the case $p=0$ do both approximations agree. 

In the special case \eqref{BR} it is easy to prove integrability explicitly. Consider the following generating function and respective canonical transformation:
%--------------
\begin{subequations}
\begin{eqnarray}
F_3(p,Q,t)&=&-p\,Q+\omega t\,Q, \\
q&=&-\frac{\partial F_3}{\partial p}=Q,\label{qQ} \\
P&=&-\frac{\partial F_3}{\partial Q}=p-\omega t,\label{pP} \\
\cal{K}(Q,P)&=&\cal{H}+\frac{\partial F_3}{\partial t}= \cal{H}+\omega Q.\label{KH}
\end{eqnarray} \label{F3}
\end{subequations}
%---------------
The new Hamiltonian $\cal{K}$ is autonomous and the system is integrable. Moreover, a stroboscopic map in the original space $(p,q)$ can be constructed at stroboscopic instants $t_k=2\pi k/\omega$, with $k\in\mathbb{N}$, by noting that $(P_k,Q_k)=(p_k,q_k)$. Since $\cal{K}$ is a constant of motion, its contour curves will correspond to the stroboscopic map for irrational values of $B/\omega$, being $B=2\sqrt{B_0²+\Omega²}$ the characteristic frequency of the integrated motion and $\Omega=B_1-\omega/2$. By \eqref{BR}, \eqref{HqpA} and \eqref{F3} we get the contours for the R stroboscopic map, namely
%--------------
\begin{eqnarray}
\cal{K}=2B_0\sqrt{1-q_k²}\cos{p_k}-2 \left(B_1-\frac{\omega}{2} \right) q_k.
\label{Rmap}
\end{eqnarray}
%---------------
From now on we use the notation $r(t_k)=r_k$. A further interesting result for this system is the analytical correspondence between periodic orbits in the classical phase space and eigenstates in the Hilbert space \cite{AW}.

\begin{figure}[ht]
\centerline{\includegraphics[scale=0.3,angle=-90]{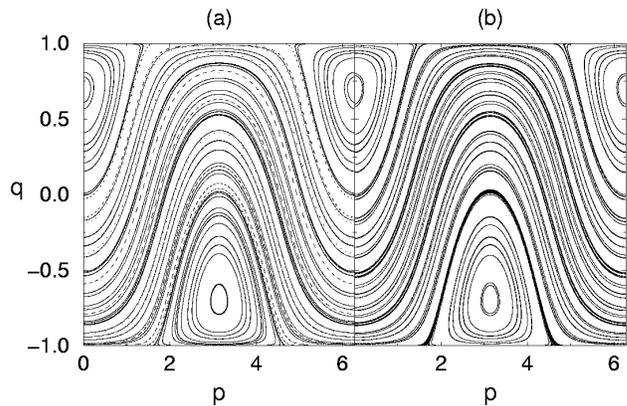}}
\caption{\small (a) Stroboscopic map for the NR system with $B_0=1.0$, $B_1=1.5$ and $\omega=3.0$ and (b) the corresponding map for the the R system with $\omega=\gamma=4.9559$ (see eq. \ref{Rmap}). This value was obtained by the fitting procedure \cite{AW}.}
\label{sec1}
\end{figure}

For the NR system \eqref{BNR}, the absence of a rotation symmetry destroys the manifest integrability of the model. Notice that the modulus of the magnetic field is no longer constant and finding a rotating frame does not suffice. Numerical calculation is now required. For convenience, hereafter we shall work with $\mathbf{B}_{NR}=-2(B_3,0,-B_2\cos{\omega t})$, which results from a rotation of the field \eqref{BNR} through the angle $\pi/2$ around the $y$-axis. We also set $B_3\rightarrow B_0$ and $B_2\rightarrow -B_1$ in order to establish contact with the field \eqref{BR}. A first remarkable result found numerically is the linear relation between the time-dependent energy, $\cal{H}(t)$, and the coordinate $q(t)$ at the stroboscopic instants \cite{AW}. 
It allows us to write $\cal{H}_k = \cal{E}(q_0,p_0)-\gamma\, q_k$ and, consequently,
%--------------
\begin{eqnarray}
\cal{E}(q_0,p_0) = 2 B_0 \sqrt{1-q_k²}\cos{p_k}-2\left(B_1-\frac{\gamma}{2} \right) q_k,
\label{NRmap}
\end{eqnarray}
%--------------
which is the equation for the NR contours. Comparison between \eqref{Rmap} and \eqref{NRmap} emphasizes the resemblance between NR and R dynamics at stroboscopic instants. It illustrates the integrability of the NR system. The qualitative agreement between NR and R maps are shown in Fig.\ref{sec1}, which was made as follows: we determined $\gamma$ numerically by fittings in the NR system and then we used a R field with $\omega=\gamma$.
%---------------------------------------------
%---------------------------------------------

The contour pattern \eqref{NRmap} ``chosen'' by the NR dynamics is not obvious. In fact it cannot be inferred directly from the Hamiltonian for the NR field. The choice of one specific contour in the set of possibilities offered by the time-dependence of the system suggests the existence of an average dynamics which underlies integrability (see \cite{AW} for details).

The integrability of the system is made evident via averaging theorem in other two limiting regimes, namely, the strong coupling $(B_0\ll\omega)$ and the resonant weak coupling ($2B_0=\omega$ and $B_1\ll\omega$) regimes. The validity of the RWA is also numerically seen to extend beyond the time scale $\cal{O}(1/B_1T)$ \cite{AW}. Thus, integrability and its consequences explain the wide validity of the RWA (see also \cite{barata00} for numerical results).

Last but not least, we describe our results for the NOT operation in the NR system. Although analytical solutions were not found in this case, we may use \eqref{Rmap} and \eqref{NRmap} to make predictions.

General principles of quantum computation require that the NOT operation be unitary \cite{gisin99}. References \cite{gisin99,buzek99} propose different constructions of the universal unitary NOT operation, which takes as input a qubit in an arbitrary state $|\psi\rangle$ and generates an output as close as possible to its orthogonal state $|\psi_{\bot}\rangle$. Since we are primarily concerned with the dynamics, we consider the special cases where a unitary operation leading from $|\psi\rangle$ to $|\psi_{\bot}\rangle$ exists.

Following the derivation of \eqref{Drho}, one may obtain $2 |\langle \psi_0|\psi_t\rangle|²=1+\mathbf{S}_0\cdot\mathbf{S}_t$. The NOT operation occurs when the dynamics produces, at the instant $t_{not}$, a state $|\psi_{not}\rangle$ orthogonal to the initial state $|\psi_0\rangle$, such that $\langle \psi_0|\psi_{not}\rangle=0$. Classically, it corresponds to $\mathbf{S}_{not}=-\mathbf{S}_0$. 

\begin{figure}[H]
\centerline{\includegraphics[scale=0.32,angle=-90]{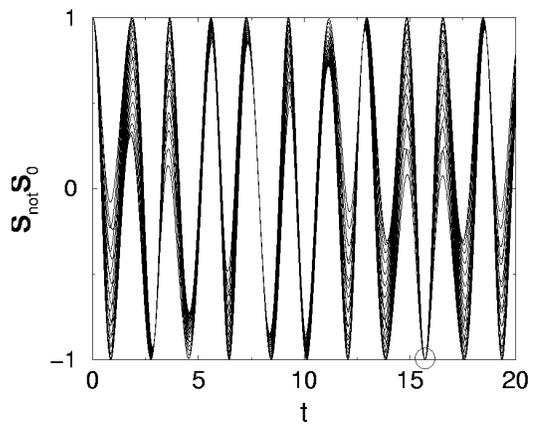}}
\caption{\small NR NOT operation for $\omega=1.0$, $\gamma=1.486$, $B_1=1.5$ and $B_0=1.279$ occurring at $t_{not}=5\pi$ for several initial conditions given by $(p_0,q_0)=\left(\frac{\pi}{2},\forall\right)$, which corresponds to a universality class of quantum states with purely complex relative phase.}
\label{ss}
\end{figure}

According to \eqref{rhoS}, the corresponding initial quantum states which allow the NOT operation are simply obtained by putting $(p_0,q_0)$ given above into
 %--------------------
\begin{eqnarray}
|\psi\rangle=\sqrt{\frac{1-q}{2}}\,|+\rangle+\sqrt{\frac{1+q}{2}}\,e^{\imath\,p}\,|-\rangle, \label{psiqp}
\end{eqnarray}
%--------------------
which emphasizes that the canonical momentum $p$ plays an important role of a relative quantum phase. In \cite{AW} we have found analytically the regimes for the realization of the unitary NOT operation on the R system. In particular, when the resonance condition $\omega²=B_0²+(B_1-\omega/2)²$ is satisfied, the NOT operation will occur periodically at $t_{not}^n=(2n+1)\pi/\omega$ for initial conditions given by arbitrary $p_0$ and $q_0=0$. Very interestingly, this regime also occurs for the NR system. Following the analogy between $\gamma$ and $\omega$ given by \eqref{Rmap} and \eqref{NRmap} we determine numerically the parameters satisfying the required resonance and the universality class of initial states for the NOT operation in the NR system (see Fig.\ref{ss}).

The predictions of NOT operations in the NR system were based on comparisons with stroboscopic maps of the R system. These classical mathematical tools, based on the concept of trajectories flowing in the phase space, were crucial in the determination of the connection between the R frequency $\omega$ and its NR counterpart $\gamma$. It stresses the usefulness of the classical gyromagnet picture, being an example of classical analysis allowing quantum predictions.

R.M.A. thanks financial support from FAPESP (Fundação de Amparo à Pesquisa do Estado de São Paulo) under grant 02/10442-6. W.F.W. thanks H.R. Jauslin and D. Sugny for remarks and suggestions.

%---------------------------------------------
%---------------------------------------------

%%%%%%%%%%%%%%%%%%%%%%%%%%%%%%%%%%%%%%%%%%%%%%%%%%%%%%%%%%

%%%%%%%%%%%%%%%%%%%%%%%%%%%%%%%%%%%%%%%%%%%%%%%%%%%%%%%%%%

\end{document}